\documentclass[twocolumn,showpacs,preprintnumbers,amsmath,amssymb,prl]{revtex4}

\usepackage{graphicx}% Include figure files
\usepackage{bm}% bold math

\begin{document}

\title{Approximate direct
correlation function for multi--Yukawa hard--core systems}

\author{E. E. Tareyeva}
\affiliation{Institute for High Pressure Physics, Russian Academy
of Sciences, Troitsk 142190, Moscow, Russia}
\author{V. N. Ryzhov}
\affiliation{Institute for High Pressure Physics, Russian Academy
of Sciences, Troitsk 142190, Moscow, Russia \\ Moscow Institute of
Physics and Technology, Dolgoprudny, Moscow Region 141700, Russia}

\date{\today}

\begin{abstract}
Simple closed analytical expression for approximate direct
correlation function (DCF) for multi--Yukawa hard--core system of
particles is presented. The obtained DCF is a solution of the
Ornstein--Zernike equation with multi--Yukawa closure valid in the
linear approximation in the potential. This approximation includes
linear corrections to the hard-- sphere DCF inside the core
radius.
\end{abstract}

 \maketitle

Despite of a lot of results on simple liquids with spherical
potentials obtained during last decades it is always interesting
and useful to complement real and computer experiments and
semiphenomenological studies by analytical calculations based on
the first principles of statistical mechanics. The advantage of
such calculations consists not only in  the fact that they are
simpler to use, but they can also give fruitful physical insights
to solving more complicated problems. The direct correlation
function (DCF) is an important source of different kinds of
information that can be derived from a microscopic theory based on
statistical mechanics principles \cite{Hans,Bal}. In particular,
DCF is very useful for study structural aspects  of liquids as
well as their thermodynamic aspects. The first analytical theories
for thermodynamic properties were the perturbation theories based
on the classical paper \cite{BaHe}. Integral equations theories
based on the Ornstein-Zernike (OZ) equation usually are considered
as nonanalytical theories needing numerical methods to proceed
(see, however, the papers \cite{Chen,Hlush,Ginoza}). In the case
of square-well fluids there exists an analytical theory \cite{TL}
that combines perturbation theory with the mean spherical
approximation. The square-well systems are rather fruitfully
investigated \cite{S2005,S2013} by using the semi-analytic
methodology -- rational-function approximation for  the
pair-correlation function \cite{Santos}. Some analytic results for
square--well fluid were obtained in our papers
\cite{PRSW,PRSW18,PRSW19,PRSW20}.

The most popular analytical approximation for the direct
correlation function $C(r)$ for systems of particles with hard
--core potentials is based on the so called mean spherical
approximation (MSA) \cite{Hans,Bal,PerLeb}. In MSA: $g(r)=0$ for
$r\leq 1$ (inside the core), and $C(r) = - \beta \phi (r)$ for
$r>1$ (the linear term of the Percus-Yevick approximation for
$C(r)$). Here $\beta $ is the inverse temperature and $\phi (r)$
the interparticle potential outside the core. The approximation
for $C(R)$ similar in spirit to MSA was proposed by Lovett
\cite{Lovett}:
\begin{eqnarray}
\label{Lovett}
 C(r)=C_{HS}(r),\hspace{3mm} r\leq 1
\\
\nonumber C(r) = - \beta \phi (r), \hspace{3mm} r>1,
\end{eqnarray}
where $C_{HS}$ is the DCF for hard--sphere system
 \cite{Baxt,Werth,Thiele}.
It is the approximation (\ref{Lovett}) that we have used in papers
\cite{PRSW,PRSW18,PRSW19,PRSW20}.

In the present paper we propose an approach initiated by the
so-called generalized mean spherical approximation (GMSA)
\cite{Hans,HoBl1} based on the works by Baxter \cite{Baxt,Baxt2}.
It is known that the direct use of GMSA meets serious technical
problems of such extend that the results can be presented only in
the form of long tables. So, we propose to simplify the approach
and to use linearized GMSA (LGMSA). We construct the DCF  in such
a way that outside the core it is equal to $-\beta \phi (r)$ as in
MSA and the whole function $C(r)$ satisfies the OZ equation in the
approximation linear in $\beta \phi $. This means that we obtain
explicitly the linear corrections to $C_{HS}(r)$ inside the core
radius. It seems that such an approximation is rather
self--consistent and to some extend even more adequate than GMSA.
It is nessessary to emphasize that the LGMSA enables us to obtain
closed ready for immediate use analytical expression for DCF in
the important case of arbitrary number of terms in the
multi-Yukawa potential. This means that the result can be used for
almost every (standard) spherical potential.

%\section{Linearized generalized mean spherical approximation (LGMSA).}

The direct correlation function $C(r)$ is connected with the pair
ceorrelation function $g(r)$ through well known OZ relation
\cite{Hans,Bal,OZ}. In the case of one--component fluid with
spherically symmetric interaction the OZ relation has the form:
\begin{equation}
h(r) = C(r) + \rho \int d{\vec r'} c(|{\vec r'}|) h(|{\vec r} -
{\vec r'}|), \label{OZ}
\end{equation}
where $h(r) = g(r) - 1$, $\rho$ is the particle number density. If
the particles of the system have a hard impenetrable core of the
radius $r=1$ then
$$h(r) = -1,  r<1.$$
If a closure is added to the Eq. (\ref{OZ}), for example, if one
proposes the form of $C(r)$ for $r>1$, then the relation
(\ref{OZ}) becomes an equation. In particular we can put $C(r) = -
\beta \phi (r)$ for $r>1$
 (see \cite{PerLeb}). The OZ relation with this closure usually is
called the Percus-Yevick equation for $C(r)$. It is this equation
for hard-sphere systems ($C(r)=0$ for $r>1$) that was solved in
the papers \cite{Baxt,Werth,Thiele} where the following solution
was obtained (we will use it in what follows):

\begin{equation}
C_{HS}(r) = -\lambda _{1} - 6 \eta \lambda _{2} r -\frac{1}{2}\eta
\lambda _{1} r^{3}, \label{CHS}
\end{equation}

where
$$\lambda _{1} = \frac{(1+2\eta)^2}{(1-\eta)^4},$$
$$\lambda _{2} = - \frac{(1+(1/2)\eta)^2}{(1-\eta)^4},$$
$\eta = \pi \rho /6$.

We base our approach on the classical works by Baxter
\cite{Baxt,Baxt2} and on the papers where the Baxter's method was
used for systems with the multi-Yukawa hard-core potential
\cite{HoBl1,HoBl2,HenB,Chen,Hlush,Ginoza,Cluster}. It is worth
noting that up to now only in the case of one Yukawa term
\cite{EOS} the explicit results were obtained.

Let us recall briefly the Baxter's method. In \cite{Baxt} it was
shown that it is possible to obtain the Wiener--Hopf factorization
for the function
$$\tilde A(k)=1- \rho \tilde C(k),$$
that is it can be presented in the form
$$\tilde A(k) = \tilde Q(k) \tilde Q(-k),$$
while the function $1-\tilde Q(k)$ is Fourier integrable along the
real axis and one can define a function $Q(r)$ as
$$2\pi \rho Q(r) = \frac{1}{2\pi }\int_{-\infty}^{\infty} exp(-ikr) [1-
\tilde Q (k)]dk.$$
$$\tilde Q(k) = 1 - 2 \pi \rho \int_0^R e^{ikr}Q(r)dr.$$
The function $Q(r)$ is a real continuous function at $r>0$ and
$Q(r)=0$ for $r<0$. If $C(r)=0$ at $r>R$, then $Q(r)=0$ at $r>R$,
too. Using this function Baxter transformed the original OZ
equation into following two equations
\begin{equation}
r C(r) = - Q'(r) + 12 \eta \int _{0}^{R}dt Q'(t) Q(t-r)
\label{Ba1}
\end{equation}
\begin{equation}
r h(r) = - Q'(r) + 12 \eta \int _{0}^{R}dt (r-t) h(|r-t|) Q(t-r)
\label{Ba2}
\end{equation}
Here $Q'(r)$ means the derivative of $Q(r)$ with respect to $r$.
One great advantage of the Eqs.(\ref{Ba1}), (\ref{Ba2}) in
comparison with the initial OZ equation is that they are
one-dimensional. The second very great advantage is that now the
special hard-core condition, namely, $h(r)=-1$ for $r<1$, can be
utilized in a very profitable way. For the system of hard spheres,
using the closure $C(r)=0$ for $r>1$ (i.e. putting $R=1$), we
obtain from (\ref{Ba2}):
\begin{equation}
Q_{HS}(r) = \frac{1}{2}a_{HS}(r^{2} -1) +b_{HS}(r-1) \label{QHS}
\end{equation}
with
\begin{equation}
a_{HS} = \frac{1+2 \eta}{(1-\eta)^{2}}, \hspace{12mm} b_{HS} =
-\frac {3 \eta}{2(1-\eta)^{2}}, \label{ABHS}
\end{equation}
and from (\ref{Ba1}) the form (\ref{CHS}).

The Baxter's method was applied in the paper \cite{HoBl1} to the
system of particles with the hard--core multi--Yukawa potential
\begin{eqnarray}
\phi (r) &=& \infty, \hspace{3mm} r\leq 1 \nonumber
\\
\phi (r) &=&
 \sum_{i=1}^{n} \frac{\epsilon _i}{r}e^{-z_i(r-1)}, \hspace{3mm} r>1,
\label{poten}
\end{eqnarray}

In this case the Baxter form of the OZ equation reads: $h(r)=-1$
for $r<1$ and
$$C(r)=-\beta \sum_{1}^{n} \frac{\epsilon  _i}{r}e^{-z_i(r-1)}\equiv
\sum_{1}^{n} \frac{K_i}{r}e^{-z_i(r-1)},$$ for $r>1$. This
approximation was named generalized mean-spherical approximation
(GMSA) \cite{Hans,HoBl1}. In this case the hard-core condition for
$h(r)$ again occurs to be very advantageous so that it makes
possible to obtain the general form of the Baxter function $Q(r)$.
The coefficients depend upon the temperature and the density and
can be obtained in principle after substituting the proposed form
of $Q(r)$ in the equations (\ref{Ba1}) and  (\ref{Ba2}). Finally
one obtains a very complicated system of $2n+2$ algebraic
equations. Only two of them are linear. This system can be solved
numerically. It was only in the case of one-Yukawa potential that
the nonlinear equations are quartic, what allows to obtain some
physical results\cite{EOS}.

We propose to exploit the linear approximation of the theory
\cite{HoBl1}. Our aim is to obtain the corrections to $C_{HS}(r)$
at $(r<1)$ linear in $-\beta \phi (r)$ . With these corrections
the whole function $C(r)$ must satisfy the OZ equation in this
linear approximation. This means, in particular, that the first of
Baxter's equations (\ref{Ba1}) now has the form ($r<1$):
\begin{eqnarray}
&&rC_{HS}(r) + r \delta C(r) = -Q'_{HS} (r)- \delta Q'(r)+\nonumber\\
&+& 12 \eta \int _{0}^{1}dt Q_{HS}'(t) Q_{HS}(t-r) +\nonumber\\
&+& 12 \eta \int _{0}^{1}dt Q_{HS}'(t) \delta Q(t-r) +\nonumber\\
&+& 12 \eta \int _{r}^{1+r}dt \delta Q'(t) Q_{HS}(t-r).
\label{LBa1}
\end{eqnarray}

Following \cite{HoBl1} ,  it is easy to show that $\delta Q(r)$
contains the following terms:
\begin{eqnarray}
\delta Q(r) &=& \frac{1}{2} a_{1} (r^{2}-1)+b_{1}(r-1)
+\nonumber\\
&+&\sum_{i=1}^{n}c_{i}(e^{-z_{i}r} - e^{-z_{i})}+
\sum_{i=1}^{n}d_{i}e^{-z_{i}r} \label{delqin}
\end{eqnarray}
if $r\leq1$ and
\begin{equation}
\delta Q(r) = \sum_{i=1}^{n}d_{i}e^{-z_{i}r} \label{delqout}
\end{equation}
if $r>1$.

The coefficients $a_{1}, b_{1}, c_{i}, d_{i}$ can be obtained from
the Eq.(\ref{Ba2}). For $r<1$ we have
\begin{eqnarray}
\label{LBa2} &-&r = -Q'_{HS}(r) - \delta Q'(r) -12 \eta
\int_{0}^{1}(r-t)Q(t) dt-
\nonumber\\
 &-&12 \eta \int_{1}^{r+1} dt (r-t) \delta Q(t)-\nonumber\\
 &-&12 \eta
\int_{1+r}^{\infty}dt (r-t) \delta Q(t) h(|r-t|).
\end{eqnarray}

To obtain the function $h(r)$ for $r>1$  one can take the Laplace
transform of the linearized equation (\ref{Ba2}) for $r>1$ and
solve the obtained equation for
$$q(s)=\int_{0}^{\infty} dt Q(t) e^{-st}.$$

Now Eq. (\ref{LBa2}) becomes closed and after integrations we
obtain the following system of $2n+2$ linear algebraic equations:
\begin{eqnarray}
b_{1} &=& -\frac{3 \eta}{2} a_{1} - 2\eta b_{1} +\nonumber\\
&+&12 \eta \sum_{i=1}^{n}
\left[c_{i}\left[\frac{1}{z_{i}^{2}}-\left(\frac{1+z_{i}}{z_{i}^{2}}
+\frac{1}{2}\right)e^{-z_{i}}\right] + \right. \nonumber\\
&+&\left.\frac{d_{i}}{z_{i}^{2}}\right], \label{eqb}
\end{eqnarray}
\begin{eqnarray}
1-a_{1} &=& -4 \eta a_{1} - 6 \eta b_{1} +\nonumber\\
&+&12 \eta \sum_{i=1}^{n}
\left[c_{i}\left[\frac{1}{z_{i}}-\frac{1+z_{i}}{z_{i}}
e^{-z_{i}}\right] +\frac{d_{i}}{z_{i}}\right], \label{eqa}
\end{eqnarray}
\begin{equation}
K_{i}e^{z_{i}} = z_{i}d_{i}[1-12 \eta q(z_{i}] \hspace{5mm}
(i=1,...,n), \label{eqd}
\end{equation}
\begin{equation}
c_{i}+d_{i}=12 \eta [\sigma (z_{i}) (c_{i} + d_{i}) - \tau
(z_{i})c_{i}e^{-z_{i}}] \hspace{2mm}(i=1,...,n). \label{eqc}
\end{equation}

In the frames of our linear approximation it is sufficient to use
for $q(s)$ the expression for hard sphere system:
\begin{equation}
q(s) = \sigma (s) - \tau (s) e^{-s}, \label{eqq}
\end{equation}
\begin{eqnarray}
\sigma (s) &=& a_{HS} \left(\frac{1}{s^{3}} -\frac{1}{2 s}\right)
+b_{HS} (\frac {1}{s^{2}} - \frac{1}{s}), \nonumber\\
\tau  (s) &=& a_{HS} \left(\frac{1}{s^{3}} +
\frac{1}{s^{2}}\right) +b_{HS} \frac {1}{s^{2}}. \label{eqs}
\end{eqnarray}

The important property of derived system is the fact that each
pair of equations (\ref{eqd})-(\ref{eqc}) for a fixed $i$ does not
depend on the other equations, so it can be easially solved to
obtain pairs $d_{i},c_{i}$. Substituting the results in the linear
equations (\ref{eqb})-(\ref{eqa}) we can obtain the corrections to
$a_{HS}$ and $b_{HS}$.

Let us write the final result for DCF for n-Yukawa potential
(\ref{poten}). We have
$$C_(r)=C_{HS}(r)+\delta C(r), r\le 1, ~~~$$
$$C(r)= \sum_{i=1}^{n} \frac{K_i}{r}e^{-z_i(r-1)},
r>1,
$$
where $K_i=-\beta \epsilon_{i} $. Now
\begin{eqnarray}\label{deltaC}
-r\delta C(r)= \lambda'_1 r +6 \eta (\lambda '_2 + \lambda
 ''_2)r^2+\frac{1}{2}\lambda '_1 r^4+
 \\\nonumber
 \sum_{i=1}^{n}\mu _i \sqrt{z_i} (1 - e^{-z_i r})+
 \sum_{i=1}^{n}\nu _i\left[cosh(z_i r) - 1\right],
\end{eqnarray}

 Here
\begin{eqnarray}
 \lambda '_1&=&2 \frac{1+2\eta }{(1-\eta )^2} \sum_{i=1}^{n}K_i
 \frac{e^{z_i}}{z_i} \frac{1}{V^{2}(z_i,\eta )}\times \nonumber\\
 &\times&\left[72 \eta ^2 A(z_i,\eta
 ) -12 \eta  (1+2 \eta ) B(z_i,\eta )\right],
\label{lam1}
 \end{eqnarray}

\begin{eqnarray}
 \lambda '_2&=&- \frac{2+\eta }{(1-\eta )^2} \sum_{i=1}^{n}K_i
 \frac{e^{z_i}}{z_i} \frac{1}{V^{2}(z_i,\eta )}\times\nonumber\\
 &\times&\left[12 \eta (1+2 \eta )
 A(z_i,\eta ) -6 \eta  (2+ \eta ) B(z_i,\eta )\right],
 \label{lam2}
 \end{eqnarray}

\begin{equation}
 \lambda ''_2= 2 (1+2 \eta ) \sum_{i=1}^{n}K_i
 \frac{e^{z_i}}{z_i} \frac{U(z_i,\eta )}{V^{2}(z_i,\eta )},
 \label{lam3}
 \end{equation}
 $$A(z, \eta ) = \frac{1}{z^2}(V(z,\eta ) - U(z, \eta) ) +
 (\frac{1+z}{z^2}+\frac{1}{2}) e^{-z} U(z,\eta ),$$
 $$B(z, \eta ) = \frac{1}{z}(V(z,\eta ) - U(z, \eta )) +
 \frac{1+z}{z} e^{-z} U(z,\eta ),$$
 $$U(z, \eta) = 1- \frac{2 \eta }{z^3} (z^3 -3 z^2 +6) + \frac{\eta
 ^2}{z^3} (z^3 -6 z^2 + 18 z - 24),$$
 $$V(z,\eta ) - U(z,\eta )= \frac{e^{-z}}{z^3} (12 \eta  (1+z) + 6 \eta ^2
 (z+4)),$$
$$\mu _i
 = 24 \eta
 K_i \frac{1}{z_i V(z_i,\eta)}\left[(1 + 2 \eta ) \frac{1+z_i}{z_{i}^3}
   - \frac{3 \eta }{2} \frac{1}{z_{i}^2}\right]
$$
\begin{eqnarray}
\nu_i  &=& 24 \eta
    K_i \frac{1}{z_i V^2(z_i, \eta)}\left[(1 + 2 \eta ) \frac{1+z_i}{z_{i}^3}
   - \frac{3 \eta }{2} \frac{1}{z_{i}^2}\right]\times\nonumber\\
   &\times&    [(V(z_i,\eta) - U(z_i,\eta)].\nonumber
\end{eqnarray}

Using DCF one can obtain the equation of state from the well-known
relation
 \cite{Hans,Bal}
\begin{equation}
(k_B T)^{-1} \left(\frac{\partial P}{\partial \rho }\right)_{T} =
1-\rho \int d{\bf r} C(r) = 1 - \rho {\tilde C}_0. \label{main}
\end{equation}

As an illustration of the obtained results let us consider the
case of two-Yukawa potential, which can be used as a qualitative
approximation for the Lennard-Jones one \cite{Bob}:
\begin{eqnarray}
\phi (r)&=& \infty, r\le 1, \nonumber\\
\phi (r)&=&  \frac{\epsilon
}{r}\left[e^{-z_1(r-1)}-e^{-z_2(r-1)}\right],
 r>1. \label{2Ypot}
\end{eqnarray}
where $z_1=14.735, z_2=2.68$. Using Eqs. (\ref{main}) and
(\ref{2Ypot}), one can estimate, for example, the critical point
of the two-Yukawa system. Solving the equations
$$\frac{\partial {P}}{\partial \eta }=0;~~~\frac{\partial^2 {P}}{\partial
   \eta ^2 }=0,$$
one can obtain the critical point of the system:
$$\eta _c = 0.18771  ,~~~~ T_c/\epsilon = 0.79795.$$

%The case of two-Yukawa potential \cite{Bob} that imitates
%Lennard-Jones one was considered in our paper \cite{TMF}. We
%obtained for the critical point (in LJ units) in LGMSA:
%$$\rho _
%c=0.3585  , \hspace{4mm} T_c=1.6757,$$ while the use of the
%approximate form (\ref{Lovett}) gives
%$$\rho _c=0.2457  , \hspace{4mm} T_c=1.0401.$$
%In \cite{TMF} the supercritical region was considered, this
%problem being to-day in the focus of investigations (see, e.g.
%\cite{UFN}) due to increasing of obtainable experimental ranges
%and to technological aspects, too \cite{kideb}.

To summarize, we propose an approximate simple closed analytical
form for DCF of hard-core multi-Yukawa systems. The obtained DCF
inside the core as well as outside the core satisfies OZ equation
in the linear in $\beta \phi(r) $ approximation. The new
approximations obtained in the present work allow to get the van
der Waals-like model which permits a purely analytical study of
fluid properties including the equation of state, phase behavior
and supercritical fluctuations. These results can be used in the
investigation of the behavior of the system in the vicinity of the
critical point, in particular, the properties of the Widom line
\cite{pnas2005,poole,fr_st,mcm_st,bryk2,jpcb,br_jcp,br_ufn,gallo_jcp,riem,may,paola,paola1,wePRE2015,weSciRep2015},
which is widely studied now because of its importance for
understanding the properties of the supercritical fluids.

\bigskip

\begin{acknowledgments}
The work was supported by Russian Science Foundation (Grant No
14-22-00093).
\end{acknowledgments}


\begin{thebibliography}{99}

\bibitem{Hans}
 J. P. Hansen and I. R. McDonald, {\it Theory of Simple Liquids}
(Academic, New York, 1986).

\bibitem{Bal}
R. Balescu, {\it Equilibrium and Nonequilibrium Statistical
Mechanics} (Wiley, New York, 1975).

\bibitem{BaHe} J.A.Barker, and D.Henderson, J.Chem.Phys. {\bf 47}, 2856
(1967).
\bibitem{Chen} M.Chen, J.Math.Phys. {\bf 27}, 1852 (1985).

\bibitem{Hlush} S.Hlushak, A.Trokhymchuk, S.Sokolowski, J.Chem.Phys. {\bf
130}, 234511 (2009).

\bibitem{Ginoza} M.Ginoza, Mol.Phys. {\bf 71}, 145 (1990); {\it ibid} {\bf
90},373 (1997); J.Phys.Condens.Matter {\bf 6}, 1439 (1994).

\bibitem{TL} Y.Tang and B.C.-Y.Lu, J.Chem.Phys. {\bf 100}, 3079, 6665
(1994).

\bibitem{S2005}
J.Largo, J.R.Solana, S.B.Yuste, and A.Santos, J.Chem.Phys., {\bf
122}, 084510 (2005).

\bibitem{S2013} A.Santos, S.B.Yuste, M.L\'opez de Haro, and P.Orea,
J.Chem.Phys. {\bf 139}, 074505 (2013).

\bibitem{Santos}
S.B.Yuste, and A.Santos, J.Chem.Phys., {\bf 101}, 2355 (1994).

\bibitem{PRSW} V.V.Brazhkin, Yu.D.Fomin, V.N.Ryzhov, E.E.Tareyeva, and
E.N.Tziok, Phys.Rev.E {\bf 89}, 042136 (2014).

\bibitem{PRSW18} V.N.Ryzhov, and E.E.Tareyeva, Physica A {\bf 314},
396 (2002).

\bibitem{PRSW19} V.N.Ryzhov, and E.E.Tareyeva,
Theor. Math. Phys. {\bf 130}, 101 (2002).

\bibitem{PRSW20}  V.N.Ryzhov, E.E.Tareyeva, and Yu.D.Fomin,
Theor. Math. Phys. {\bf 167}, 645 (2011).

\bibitem{PerLeb} J.K.Percus and G.Yevick, Phys.Rev. {\bf 136}, 290
(1964); J.L.Lebowitz and J.K.Percus, Phys.Rev. {\bf 144}, 251
(1965).

\bibitem{Lovett}
 R. Lovett, J. Chem. Phys. 66, 5 (1977).

\bibitem{OZ} L.S.Ornstein, and F.Zernike, Proc. Acad. Sci. Amsterdam
{\bf 17}, 793 (1914).

\bibitem{Baxt}
Baxter R.J., Austral.J.Phys.,{\bf 21}, 563 (1968).

\bibitem{Werth}Wertheim M.S., Phys.Rev.Lett., {\bf 10}, 321 (1963);
J.Math.Phys. {\bf 5}, 643 (1964).

\bibitem{Thiele} Thiele E., J. Chem.Phys., {\bf 39}, 474 (1963).

\bibitem{HoBl1}
 J. S. Hoye and L. Blum, J. Stat. Phys. {\bf 16}, 399 (1977).

\bibitem{Baxt2} Baxter R.J., J.Chem.Phys., {\bf 52}, 4559 (1970).

\bibitem{HoBl2}
 L. Blum and J.S.Hoye, J. Stat. Phys. {\bf 19}, 317 (1978).

\bibitem{HenB} D. Henderson, L.Blum, J.P.Novoryta, J.Chem.Phys. {\bf 102},
4973 (1995).

\bibitem{Cluster} Yun Liu, Wei-Ren Chen, Sow-Hsin Chen , J.Chem.Phys.,
{\bf 122}, 044507 (2005).

\bibitem{EOS} J.H.Herrera, H.Ruiz-Estrada, L.Blum, J.Chem.Phys.{\bf 104},
6327 (1996).

\bibitem{Bob} M.Bahaa Khedr, S.M.Osman, M.S.Al Busaidi, Physics and
   Chemistry of Liquids. {\bf 47}, 237-249 (2009).

%\bibitem{TMF}TMF

\bibitem{pnas2005} L. Xu, P. Kumar, S.V. Buldyrev, S.-H. Chen , P.H. Poole, F. Sciortino, and
 H.E. Stanley, Proc. Natl. Acad. Sci. U.S.A. {\bf 102}, 16558 (2005).
 \bibitem{poole} P.H. Poole, S.R. Becker, F. Sciortino, and F.W. Starr, J. Phys. Chem. B
{\bf 115}, 14176 (2011).

\bibitem{fr_st} G. Franzese and H.E. Stanley, J. Phys.: Condens. Matter {\bf 19}, 205126 (2007).

\bibitem{mcm_st} P.F. McMillan and E.H. Stanley, Nat. Phys. {\bf 6}, 479 (2010).

%\bibitem{bryk1}  G.G.Simeoni, T. Bryk, F.A. Gorelli, M. Krisch, G. Ruocco,
%M. Santoro, and T. Scopigno, Nat. Phys. {\bf 6}, 503 (2010).

\bibitem{bryk2} F.A. Gorelli, T. Bryk, M. Krisch, G. Ruocco, M. Santoro, and
T. Scopigno, Sci. Rep. {\bf 3}, 1203 (2013).

\bibitem{jpcb} V.V. Brazhkin, Yu.D. Fomin, A.G. Lyapin, V.N. Ryzhov, and E.N. Tsiok,
J. Phys. Chem. B {\bf 115}, 14112 (2011).

\bibitem{br_jcp}  V.V. Brazhkin and V.N. Ryzhov, J. Chem. Phys. {\bf 135}, 084503 (2011).

\bibitem{br_ufn}  V.V. Brazhkin, A.G. Lyapin, V.N. Ryzhov, K. Trachenko,
 Yu.D. Fomin, and E.N. Tsiok, Usp. Fiz. Nauk {\bf 182}, 1137 (2012)
 [Sov. Phys. Usp. {\bf 55}, 1061 (2012)].

\bibitem{gallo_jcp} P. Gallo, D. Corradini, and M. Rovere, J.
Chem. Phys. {\bf 139}, 204503 (2013).

\bibitem{riem}  G. Ruppeiner, A. Sahay, T. Sarkar, and G. Sengupta,
Phys. Rev. E {\bf 86}, 052103 (2012).

\bibitem{may}  H.-O. May and P. Mausbach, Phys. Rev. E {\bf 85}, 031201 (2012).

%\bibitem{UFN} Â.Â.Áðàæêèí, À.Ã.Ëÿïèí, Â.Í.Ðûæîâ, Ê.Òðà÷åíêî, Þ.Ä.Ôîìèí,
%Å.Í.Öèîê. ÓÔÍ, {\bf 182}, 1137 (2012).

%\bibitem{kideb}
%Kiran E, Debenedetti P G, Peters C J (Eds) {\it Supercritical
%Fluids. Fundamentals and Applications} (Dordrecht: Kluwer Acad.
%Publ., 2000)

\bibitem{paola} P. Gallo, D. Corradini and M. Rovere,
Nature Communications  {\bf 5}, 5806 (2014).

\bibitem{paola1} D. Corradini, M. Rovere, and P. Gallo, J. Chem.
Phys. {\bf 143}, 114502 (2015).

\bibitem{wePRE2015} Yu. D. Fomin, V. N. Ryzhov, E. N. Tsiok and V. V.
Brazhkin, Phys. Rev. E {\bf 91}, 022111 (2015).

\bibitem{weSciRep2015} Yu. D. Fomin, V. N. Ryzhov,
E. N. Tsiok and V. V. Brazhkin,  Sci. Rep. {\bf 5}, 14234 (2015).



\end{thebibliography}

\end{document}